# HEAX: An Architecture for Computing on Encrypted Data

M. Sadegh Riazi
UC San Diego
mriazi@ucsd.edu

Kim Laine
Microsoft Research
kim.laine@microsoft.com

Blake Pelton
Microsoft
blakep@microsoft.com

Wei Dai
Microsoft Research
wei.dai@microsoft.com

## Abstract

With the rapid increase in cloud computing, concerns surrounding data privacy, security, and confidentiality also have been increased significantly. Not only cloud providers are susceptible to internal and external hacks, but also in some scenarios, data owners cannot outsource the computation due to privacy laws such as GDPR, HIPAA, or CCPA. Fully Homomorphic Encryption (FHE) is a groundbreaking invention in cryptography that, unlike traditional cryptosystems, enables computation on encrypted data without ever decrypting it. However, the most critical obstacle in deploying FHE at large-scale is the enormous computation overhead.

In this paper, we present HEAX, a novel hardware architecture for FHE that achieves unprecedented performance improvements. HEAX leverages multiple levels of parallelism, ranging from ciphertext-level to fine-grained modular arithmetic level. Our first contribution is a new highly-parallelizable architecture for number-theoretic transform (NTT) which can be of independent interest as NTT is frequently used in many lattice-based cryptography systems. Building on top of NTT engine, we design a novel architecture for computation on homomorphically encrypted data. Our implementation on reconfigurable hardware demonstrates 164–268× performance improvement for a wide range of FHE parameters.

**ACM Reference Format:**
M. Sadegh Riazi, Kim Laine, Blake Pelton, and Wei Dai. 2020. HEAX: An Architecture for Computing on Encrypted Data. In *Proceedings of the Twenty-Fifth International Conference on Architectural Support for Programming Languages and Operating Systems (ASPLOS '20), March 16–20, 2020, Lausanne, Switzerland.* ACM, New York, NY, USA, 15 pages. https://doi.org/10.1145/3373376.3378523

## 1 INTRODUCTION

Cloud computing has, in a short time, fundamentally changed the economics of computing. It allows businesses to quickly and efficiently scale to almost arbitrary-sized workloads; small organizations no longer need to own, secure, and maintain their own servers. However, cloud computing comes with significant risks that have been analyzed in the literature over the last decade (see [26, 38, 57]). Specifically, many of these risks revolve around data security and privacy. For example, data in cloud storage might be exposed to both outsider and insider threats, and be prone to both intentional and unintentional misuse by the cloud provider. Recently, the European Union and the State of California have passed strong data privacy regulations. In this light, companies and organizations that possess highly private data are hesitant to migrate to the cloud, and cloud providers are facing increasing liability concerns.

To mitigate security and privacy concerns, cloud providers should keep customers' data encrypted at all times. Symmetric-key encryption schemes, such as Advanced Encryption Standard (AES) [22], allow private data to be stored securely in a public cloud indefinitely. However, unless the customers share their secret keys with the cloud, the cloud becomes merely a storage provider.

In 2009, a new class of cryptosystems, called Fully Homomorphic Encryption (FHE) [33], was introduced that allows arbitrary *computation on encrypted data*. This groundbreaking invention enables clients to encrypt data and send ciphertexts to a cloud that can evaluate functions on ciphertexts. Final and intermediate results are encrypted, and only the data owner who possesses the secret key can decrypt data, providing *end-to-end* encryption for the client.

FHE provides provable security guarantees without any trust assumptions on the cloud provider, and it can be used to enable several secure and privacy-preserving cloud-based solutions. For instance, in the context of Machine Learning as a Service (MLaaS), FHE can be used to perform oblivious neural network inference [25, 35]: clients send the encrypted version of their data, the cloud server runs ML models on the encrypted queries, and returns the result to the clients. All intermediate and final results are encrypted and can only be decrypted by the clients. Perhaps, the most critical obstacle today to deploy FHE at large-scale is the enormous computation overhead compared to a plaintext counterpart in which data is not kept confidential.

Most FHE schemes, i.e., BGV [11], BFV [31], and TFHE [18] schemes, perform exact computation on encrypted data. A recently proposed FHE scheme called CKKS [17] performs approximate computation of real numbers and supports efficient *truncation* of encrypted values. Several works [40, 43] have shown the benefits of choosing the CKKS scheme over other schemes when an approximate computation is required, e.g., in Machine Learning applications. Therefore, we focus on the CKKS scheme in this paper, even though our core modules are applicable to most of the FHE schemes.

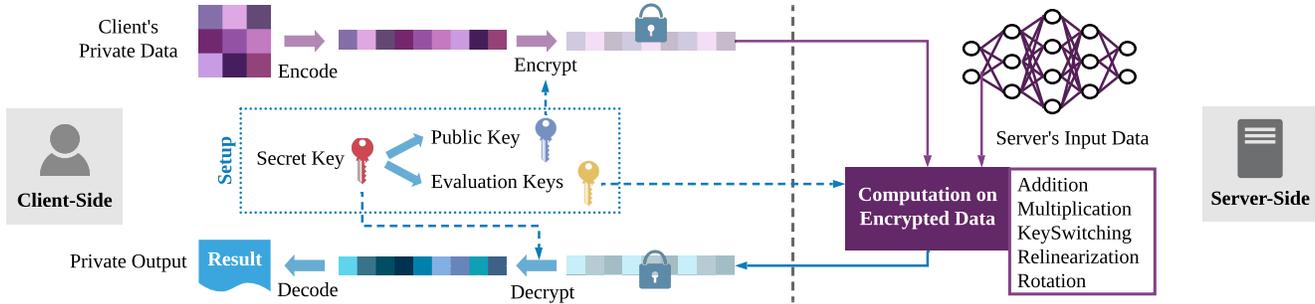

**Figure 1.** The data flow of an end-to-end encrypted computation based on homomorphic encryption.

In this paper, we introduce HEAX (stands for Homomorphic Encryption Acceleration): a novel high-performance architecture for computing on (homomorphically) encrypted data. We design several optimized core computation blocks for fast modular arithmetic and introduce a new architecture for high-throughput Number-Theoretic Transform (NTT). Building on top of the NTT module we design modules to perform high-level operations supported by FHE, thus accelerating any FHE-based privacy-preserving system.

**Prior Art and Challenges.** The *ciphertext* in FHE schemes is a set (usually a pair) of polynomials with degree $n-1$ (vectors of $n$ integers) modulo a big integer. One of the main challenges of designing an architecture for FHE is that homomorphic operations on ciphertexts involve computationally intensive modular arithmetic on big integers (with several hundred bits). These operations have convoluted data dependency among different parts of the computation, making it challenging to design a high-throughput architecture. Moreover, the degree of the underlying polynomials is enormous (in the order of several thousand). Storing the entire intermediate results on FPGA chip is prohibitive.

Prior work that propose customized hardware for non-CKKS schemes have taken one of these approaches: (i) Designing co-processors that only accelerate certain low-level ring operations [14, 19, 20, 30, 39, 61]; high-level operations are performed on the CPU-side, which makes the co-processors of limited practical use. (ii) Storing intermediate results on off-chip memory, which significantly degrades the performance [51] to the extent that it can be worse than naive software execution [53]. (iii) Designing a hardware for a fixed modest-sized parameter, e.g., $n = 2^{12}$ [54]. However, encryption parameters determine the security-level and the maximum number of consecutive multiplications that one can perform on ciphertext, both of which are application-dependent. One of our primary design goals in HEAX is to have an architecture that can be readily used for a wide range of encryption parameters. In addition, we propose several techniques to efficiently store and access data from on-chip memory and minimize (or eliminate for some parameter sets) off-chip memory accesses.

**Client-Side and Server-Side Computation.** Figure 1 illustrates the data flow and the operations involved in a typical application based on FHE. The client encrypts her data and sends the resulting ciphertext to the cloud. The cloud server performs the computation on encrypted data and sends the result back to the client. In order to perform SIMD-style operations, an *encoding* step is performed by the client to embed many numbers in a single ciphertext. Note that encoding/decoding and encryption/decryption are performed on the client-side. These operations are not computationally expensive; thus, we do not implement customized hardware for these operations. The operations that are performed by the server for *evaluating* a function on ciphertexts are computationally intensive and are the focus of this work.

**Contributions.** In what follows, we elaborate on our major contributions in more detail:

- We design a novel architecture for number-theoretic transform which is a fundamental building block – and usually the computation bottleneck – for many lattice-based cryptosystems including all FHE schemes. Our design can process arbitrary-sized polynomials with an adjustable throughput. We develop several techniques to overcome the challenges due to the complex data-dependency and convoluted access patterns within NTT.
- We introduce the *first* architecture for CKKS homomorphic encryption. We leverage multi-layer parallelism design starting from ciphertext-level to fine-grained optimized modular arithmetic engines. In contrast to the prior art for other FHE schemes, our architecture can be scaled for different FPGA chips due to its modularity. Moreover, HEAX is not custom-designed for specific FHE parameter set and can be used for a broad range of parameters.
- We provide a proof-of-concept implementation on two Intel FPGAs that represent two different classes of FPGAs in terms of available resources. We implement all high-level operations supported by CKKS and evaluate our design for three sets of FHE parameters. Our experimental results demonstrate more than *two orders of magnitude* performance improvement compared to heavily-optimized Microsoft SEAL library running on CPU.

## 2 PRELIMINARIES

**CKKS Scheme.** The homomorphic property of FHE schemes enables computation on encrypted data without the access to the decryption key. For example, adding two ciphertexts results in a ciphertext that encrypts "summation of the corresponding plaintext values". Multiplication, however, is significantly more complicated. It increases the number of polynomials in the resulting ciphertext; requiring an operation, called *relinearization*, to transform the ciphertext back to a pair of polynomials. In order to avoid the underlying plaintext values in the ciphertext to blow-up, an operation called *rescaling* is performed which divides the plaintext value by a constant number. To enable SIMD-style operations, an encoding step is performed by the client to embed many numbers in a single ciphertext. CKKS scheme supports *rotation* in which the numbers encoded in a ciphertext can be rotated.

Relinearization, rescaling, and rotation operations can be expressed as a unified operation called *Key Switching* (plus certain pre- and/or post-processing steps). Modular arithmetic operations can be computed more efficiently if ciphertext coefficients are represented in a Residue Number System (RNS). The full-RNS variant of the CKKS scheme was introduced in [16]. Another orthogonal optimization based on NTT provides a more efficient polynomial multiplication. In what follows, we provide more background on CKKS.

**Notation.** Throughout the paper, integers and real numbers are written in normal case, e.g. $q$. Polynomials and vectors are written in bold, e.g. $\mathbf{a}$. Vectors of polynomials and matrices are written in upper-case bold, e.g. $\mathbf{A}$. We use subscripts to denote the indices, e.g. $\mathbf{a}_i$ is the $i$-th polynomial or row of $\mathbf{A}$.

We assume that $n$ is a power-of-two integer and define a polynomial ring $R = \mathbb{Z}[X]/(X^n + 1)$ whose elements have degrees at most $n-1$ since $X^n = -1 \in R$. We write $R_q = R/qR$ for the residue ring of $R$ modulo an integer $q$ whose elements have coefficients in $[-\lfloor (q-1)/2 \rfloor, \lfloor q/2 \rfloor] \cap \mathbb{Z}$. In the actual computation, we represent coefficients in $[0, q-1] \cap \mathbb{Z}$. We denote by $\mathbf{u} \cdot \mathbf{v}$ the multiplication of two polynomials where the product is reduced modulo $X^n + 1$ in $R$ and further reduced modulo $q$ in $R_q$. We denote by $\langle \mathbf{u}, \mathbf{v} \rangle$ the dot product of two vectors, which gives $\sum_i u_i \cdot v_i$. We denote by $\mathbf{u} \odot \mathbf{v}$ the coefficient-wise multiplication $(u_0 \cdot v_0, u_1 \cdot v_1, \ldots)$.

For a real number $r$, $\lfloor r \rceil$ denotes the nearest integer to $r$, and $\lfloor r \rfloor$ is the largest integer smaller than or equal to $r$. For an integer $a$, $[a]_p$ denotes the reduction of $a$ modulo an integer $p$ to $[0, p-1] \cap \mathbb{Z}$. We use $\mathbf{a} \leftarrow \chi$ to denote sampling $\mathbf{a}$ according to distribution $\chi$. For a finite set $\mathbb{S}$, $U(\mathbb{S})$ denotes the uniform distribution on $\mathbb{S}$.

**Residue Number System (RNS).** There is a well-known technique to achieve asymptotic/practical improvements in polynomial arithmetic over $R_q$ with an RNS by choosing $q = \prod_{i=0}^{L} p_i$ where $p_i$'s are pair-wise coprime integers, based on the ring isomorphism $R_q \mapsto \prod_{i=0}^{L} R_{p_i}$.

**Algorithm 1** Optimized Modular Mult. | MulRed($x, y, y', p$)

**Input:** $x, y \in \mathbb{Z}_p$, $p < 2^{w-2}$, and $y' = \lfloor y \cdot 2^w / p \rfloor$
**Output:** $z \leftarrow x \cdot y \pmod{p}$
1: $z \leftarrow x \cdot y \pmod{2^w}$ ▷ the lower word of the product
2: $t \leftarrow \lfloor x \cdot y' / 2^w \rfloor$ ▷ the upper word of the product
3: $z_\epsilon \leftarrow t \cdot p \pmod{2^w}$ ▷ the lower word of the product
4: $z \leftarrow z - z_\epsilon$ ▷ single-word subtraction
5: **if** $z \geq p$ **then**
6: $\quad z \leftarrow z - p$
7: **end if**

We denote the RNS representation of an element $\mathbf{a} \in R_q$ by $\overline{\mathbf{A}} = \left( \overline{\mathbf{a}}_i = [\mathbf{a}]_{p_i} \right)_{0 \leq i \leq L} \in \prod_{i=0}^{L} R_{p_i}$. The inverse mapping is defined based on the formula $\mathbf{a} = \sum_{i=0}^{L} \overline{\mathbf{a}}_i \pi_i \left[ \pi_i^{-1} \right]_{p_i} \pmod{q}$, where $\pi_i = \frac{q}{p_i}$. Multiplications or additions in $R_q$, denoted by $\mathbf{c} = \text{Func}(\mathbf{a}, \mathbf{b})$, can be performed on their RNS representation: $\overline{\mathbf{c}}_i = \text{Func}(\overline{\mathbf{a}}_i, \overline{\mathbf{b}}_i)$ in $R_{p_i}$ (in parallel), $i = 0, 1, \ldots, L$.

**Gadget Decomposition.** Let $\mathbf{g} \in \mathbb{Z}^d$ be a gadget vector and $q$ an integer. The gadget decomposition, denoted by $\mathbf{g}^{-1}$, is a function from $R_q$ to $R^d$ which transforms an element $\mathbf{a} \in R_q$ into $\mathbf{A} \in R^d$, a vector of small polynomials such that $\mathbf{a} = \langle \mathbf{g}, \mathbf{A} \rangle \pmod{q}$. We integrate the RNS-friendly gadget decomposition from [8, 36].

**CKKS Subroutines.** We briefly review relevant subroutines:
• CKKS.Setup($\lambda$): For a security parameter $\lambda$, set a ring size $n$, a ciphertext modulus $q$, a special modulus $p$ coprime to $q$, and a key distribution $\chi$ and an error distribution $\Omega$ over $R$.
• CKKS.SymEnc(**m**, sk): Let $\mathbf{m} \in R$ be a given plaintext and sk $= \mathbf{s} \in R_{qp}$ be a secret key. Sample $\mathbf{a} \leftarrow U(R_{qp})$ and $\mathbf{e} \leftarrow \Omega$, compute $\mathbf{b} = -\mathbf{a} \cdot \mathbf{s} + \mathbf{e} \in R_{qp}$, and return the ciphertext ct $= (\mathbf{c}_0, \mathbf{c}_1) = (\mathbf{b}, \mathbf{a})$.
• CKKS.KeyGen(): Sample $\mathbf{s} \leftarrow \chi$. Return a secret key sk $= \mathbf{s}$ and a public key pk $=$ SymEnc(0, sk).
• CKKS.KskGen(sk', sk): Let sk $= \mathbf{s} \in R_{qp}$ be the generated secret key, sk' $= \mathbf{s}' \in R_{qp}$ be a different key, and a gadget vector $\mathbf{g} \in \mathbb{Z}^d$. Return a key switching key ksk $= (\mathbf{D}_0 \mid \mathbf{D}_1) \in R_{q\ell p}^{(L+2) \times 2}$, where $(\mathbf{d}_{0,i}, \mathbf{d}_{1,i}) \leftarrow$ SymEnc($g_i \cdot \mathbf{s}', \mathbf{s}$) for $i = 0, 1, \ldots, d-1$.
• CKKS.Add(ct$_0$, ct$_1$): Given ciphertexts ct$_0$, ct$_1 \in R_{q_\ell}^2$ encrypting pt$_0$, pt$_1 \in R$, generate ct' $=$ ct$_0 +$ ct$_1 \in R_{q_\ell}^2$ which is equivalent to the encryption of pt$_0 +$ pt$_1 \in R$.

Two frequently used operations in homomorphic evaluation are modular reduction and modular multiplication:
• Mod($x, p$): Used to perform modular reduction of a single-word or double-word integer [9]. For a modulus $p$ with at most $w$ bits, given an integer $x \in [0, (p-1)^2]$, precompute $u = \lfloor 2^{2w}/p \rfloor$, and compute $z = x \pmod{p}$. Mod($\mathbf{a}, p$) performs Mod($a_i, p$) for all $i = 0, 1, \ldots, n-1$.
• MulRed($x, y, y', p$): For $w$-bit words and a modulus $p < 2^{w-2}$, given $x, y \in \mathbb{Z}_p$ and precomputed $y' = \lfloor y \cdot 2^w / p \rfloor$, compute $x \cdot y \pmod{p}$ according to Algorithm 1.

## 3 MULT MODULE

In this section, we describe our proposed architectures for homomorphic multiplication.

### 3.1 Homomorphic Multiplication Algorithm

This operation is performed in RNS and NTT form. Although in general ciphertexts can have more than two polynomial components, in practice, ciphertexts are usually relinearized and the multiplication is carried out on two components as discussed next. Nevertheless, our proposed architecture is generic and supports any number of components.

• CKKS.Mul(ct$_0$, ct$_1$): Given ciphertexts ct$_0$, ct$_1 \in R_{q_\ell}^2$ encrypting pt$_0$, pt$_0 \in R$, generate ct$' \in R_{q_\ell}^3$ according to Algorithm 2 which encrypts pt$_0 \cdot$ pt$_1 \in R$.

---

**Algorithm 2** Homomorphic Mult. | CKKS.Mul(ct$_0$, ct$_1$)

**Input:** ct$_0 = (\tilde{A}_0, \tilde{A}_1)$, ct$_1 = (\tilde{B}_0, \tilde{B}_1) \in (\prod_{i=0}^{\ell} R_{p_i})^2$
**Output:** ct $= (\tilde{C}_0, \tilde{C}_1, \tilde{C}_2) \in (\prod_{i=0}^{\ell} R_{p_i})^3$
1: **for** ($i = 0;\ i \leq \ell;\ i = i + 1$) **do**
2: $\quad \tilde{c}_{0,i} \leftarrow \text{Mod}(\tilde{a}_{0,i} \odot \tilde{b}_{0,i},\ p_i)$ ▷ Dyadic Core
3: $\quad \tilde{c}_{1,i} \leftarrow \text{Mod}(\tilde{a}_{0,i} \odot \tilde{b}_{1,i} + \tilde{a}_{1,i} \odot \tilde{b}_{0,i},\ p_i)$
4: $\quad \tilde{c}_{2,i} \leftarrow \text{Mod}(\tilde{a}_{1,i} \odot \tilde{b}_{1,i},\ p_i)$
5: **end for**

---

### 3.2 HEAX Word Size and Native Operations

Microsoft SEAL library [56] is developed for x86 architectures with 64-bit native operations. However, on FPGAs, the bit-width of Digital Signal Processing (DSP) units that perform multiplication may vary, hence, it is more efficient to have a flexible bit-width for native operations. For example, the two FPGA chips that we have implemented our architecture on have 27-bit DSP units. Choosing 27-bit or 54-bit words enables us to use fewer DSPs to do the same computation. Naive construction of a 64-bit multiplier requires nine 27-bit DSPs. Whereas, a 54-bit multiplier requires only four. However, by reducing the bit-width of the RNS bases, one may need to increase the number of RNS bases; roughly speaking, by a factor of $\frac{64}{54} \approx 1.2$. In practice, small ciphertext moduli are usually less than 54 bits and thus, we do not need to increase the number of moduli.

It is worth-mentioning that leveraging more sophisticated multi-word multiplication algorithms such as Toom-Cook, one can implement 64-bit multiplication using *five* 27-bit multipliers together with more bit-level and Addition operations. Overall, by switching from 64-bit native operations to 54-bit, we observe between 1.4–2.25× *reduction in the number of DSP units* needed (depending on the HE parameters). However, to support 54-bit word size, we need to make sure that all of the ciphertext moduli ($p_i$) are (i) less than 52-bit to ensure the correctness of Algorithm 1 and (ii) congruent to 1 mod 2$n$ to support NTT as described in Section 4. We have modified the SEAL library accordingly and precomputed all of such moduli for different parameter sets.

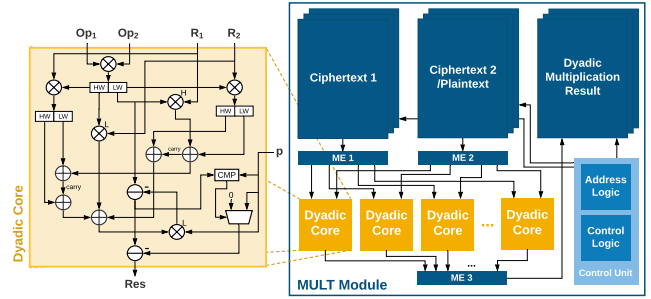

**Figure 2.** Architecture of MULT module.

### 3.3 MULT Architecture

The MULT module can process both ciphertext-ciphertext (C-C) as well as ciphertext-plaintext (C-P) homomorphic multiplications. We describe the architecture for C-C multiplication as C-P is a special case of C-C. Since ciphertexts are in NTT form by default, homomorphic multiplication is simply a series of *dyadic* products on different components.

The MULT module, as depicted in Figure 2, encompasses $nc_{\text{DYD}}$-many *Dyadic Cores*; thus, it can compute $nc_{\text{DYD}}$ dyadic multiplication at each clock cycle ($nc$ stands for number of cores). Each Dyadic core takes as input two polynomial coefficients (Op$_1$ and Op$_2$), two precomputed constant values (R$_1$ and R$_2$), and one-word prime $p$ and outputs the result.

Let us denote the number of components in ct$_0$ and ct$_1$ by $\alpha$ and $\beta$, respectively. The outcome of homomorphic multiplication is a ciphertext with $\alpha + \beta - 1$ components. Each ciphertext component is represented in a RNS form. Recall that in homomorphic multiplication (Algorithm 2), the computation can be carried out independently on each RNS basis. We leverage this property to reduce BRAM utilization. Minimum BRAM utilization is achieved by storing only one residue of one ciphertext component on FPGA chip. However, this approach significantly increases data transfer from CPU to FPGA from $(\alpha + \beta) \cdot n$ words to $(\alpha \cdot \beta + \min(\alpha, \beta)) \cdot n$ words because we need to compute all pairwise combinations of ct$_0$ and ct$_1$ components. Thus, we allocate $\alpha$-many memories of size $n$ for ct$_0$ and $\beta$-many memories for ct$_1$ to hold *one* residue of *all* ciphertext components. As a result, we achieve $O((\alpha + \beta) \cdot n)$ data transfer and BRAM consumption.

In order to fully utilize all $nc_{\text{DYD}}$ Dyadic cores – regardless of the values of $\alpha$ and $\beta$ – we read $nc_{\text{DYD}}$ coefficients from one of the polynomials of ct$_0$ and ct$_1$ at every clock cycle. However, each unit of on-chip memory, i.e., Block RAMs (BRAM), only supports one read and one write at each clock cycle. In order to read many coefficients from one polynomial at each cycle, we store each polynomial across $nc_{\text{DYD}}$-many parallel memory blocks that share common read/write address signals as depicted in Figure 2. Let us call the aggregation of one row among different BRAMs as a *memory element* (ME). Therefore, at every cycle, one memory element (ME1/ME2) is read from ct$_0$/ct$_1$ memory banks and the result (ME3) is written to a separate output memory.

# 4 NTT MODULE

NTT calculation as well as its inverse (INTT) are the most computationally intensive *low-level* operations. Polynomial multiplication is more efficiently performed by transforming polynomials and using the convolution theorem. In what follows, we provide an overview on NTT algorithm followed by our proposed architecture.

## 4.1 Algorithms

Computing $\mathbf{c} = \mathbf{a} \cdot \mathbf{b} \in R_p$ is equivalent to computing the negacyclic convolution of their coefficient vectors in $\mathbb{Z}_p^n$: $c_j = \sum_{i=0}^{j} a_i b_{j-i} - \sum_{i=j+1}^{n-1} a_i b_{j-i+n} \pmod{p}$, $j = 0, 1, \ldots, n-1$. For a large $n$ it is asymptotically better to use the convolution theorem and perform a specific form of fast Fourier transform, i.e., NTT, over a finite field. Polynomials are kept in NTT form to reduce the number of NTT/INTT conversions. Fast NTT algorithms are well studied in lattice-based cryptography. We adapt the algorithms in [44] which analyzes fast NTT algorithms and introduces specific optimizations for negacyclic convolution. For a ring degree $n$, we choose a prime number $p = 1 \bmod 2n$ such that there exists a $2n$-th primitive root of unity $\psi$, i.e., $\psi^n = -1 \bmod p$.

- $\mathrm{NTT}_p(\mathbf{a})$: Given $\mathbf{a} \in \mathbb{Z}_p^n$, compute $\tilde{\mathbf{a}} \in \mathbb{Z}_p^n$ such that $\tilde{a}_j = \sum_{i=0}^{n-1} a_i \psi^{(2i+1)j}$, according to Algorithm 3.

An important operation that is used during key switching and rescaling is called *flooring* which can be formalized as:

- $\mathrm{Floor}(\tilde{\mathbf{C}}, p)$: Given $\tilde{\mathbf{C}}$, the RNS and NTT form of $\mathbf{c} \in R_{q_\ell p}$, generate $\tilde{\mathbf{C}}'$, the RNS and NTT form of $\mathbf{c}' = \lfloor p^{-1} \cdot \mathbf{c} \rceil \in R_{q_\ell}$ according to Algorithm 4.

## 4.2 NTT Architecture

In what follows, we use the term NTT to refer to both NTT and INTT operations/modules for simplicity. At the end of this section, we discuss the differences between these two modules. As can be seen from Algorithm 5, in KeySwitch, NTT is frequently used in different parts of this algorithm. However, the number of required transformations is not consistent in different parts of the Algorithm. In order to have a fully-pipelined architecture, we allocate one NTT module per each NTT operation in Algorithm 5. However, the relative throughput-rate among different NTT instances depends on the chosen FHE parameters, which is application-dependent. As a result, we need to have a generic architecture such that the *throughput can be adjusted* as needed. This, in turn, is translated to the number of NTT *cores* that is dedicated to a given NTT module.

**NTT Core.** Figure 3 shows the internal architecture of an NTT core. Each core accepts two coefficients ($c_{\mathrm{in.a}}$ and $c_{\mathrm{in.b}}$), one twiddle factor ($w$), one precomputed value ($wp$), and a prime number ($p$) as inputs and computes two transformed coefficients as the outputs ($c_{\mathrm{out.a}}$ and $c_{\mathrm{out.b}}$). The modular arithmetic operations within NTT core are all pipelined to maximize the throughput of the overall NTT module.

---

**Algorithm 3** Number-Theoretic Transform (NTT) | $\mathrm{NTT}_p(\mathbf{a})$

**Input:** $\mathbf{a} \in \mathbb{Z}_p^n, p \equiv 1 \mod 2n$, $\mathbf{Y} \in \mathbb{Z}_p^n$ storing powers of $\psi$ in bit-reverse order, and $\mathbf{Y}' = \lfloor \mathbf{Y} \cdot 2^w / p \rfloor$.
**Output:** $\tilde{\mathbf{a}} \leftarrow \mathrm{NTT}_p(\mathbf{a})$ in bit-reverse ordering.
1: **for** ($m = 1$; $m < n$; $m = 2m$) **do**
2:   **for** ($i = 0$; $i < m$; $i$ + +) **do**
3:     **for** ($j = \frac{i \cdot n}{m}$; $j < \frac{(2i+1)n}{2m}$; $j$ + +) **do**
4:       $v = \mathrm{MulRed}(a_{j+\frac{n}{m}}, y_{m+i}, y'_{m+i}, p)$
5:       $a_{j+\frac{n}{m}} = a_j - v \pmod{p}$
6:       $a_j = a_j + v \pmod{p}$     ▷ NTT Core
7:     **end for**
8:   **end for**
9: **end for**
10: $\tilde{\mathbf{a}} \leftarrow \mathbf{a}$

---

**Algorithm 4** RNS Flooring | $\mathrm{Floor}(\tilde{\mathbf{C}}, p)$

**Input:** $\tilde{\mathbf{C}} = (\tilde{\mathbf{c}}_0, \ldots, \tilde{\mathbf{c}}_{\ell+1}) \in \mathbb{Z}_{p_0}^n \times \cdots \times \mathbb{Z}_{p_\ell}^n \times \mathbb{Z}_p^n$.
**Output:** $\tilde{\mathbf{C}}' = (\tilde{\mathbf{c}}'_0, \ldots, \tilde{\mathbf{c}}'_\ell) \in \mathbb{Z}_{p_0}^n \times \cdots \times \mathbb{Z}_{p_\ell}^n$.
1: $\mathbf{a} \leftarrow \mathrm{INTT}_p(\tilde{\mathbf{c}}_{\ell+1})$     ▷ INTT Module
2: **for** ($i = 0$; $i \leq \ell$; $i = i + 1$) **do**
3:   $\bar{\mathbf{r}} \leftarrow \mathrm{Mod}(\mathbf{a}, p_i)$
4:   $\tilde{\mathbf{r}} \leftarrow \mathrm{NTT}_{p_i}(\bar{\mathbf{r}})$     ▷ NTT Module
5:   $\tilde{\mathbf{c}}'_i \leftarrow \tilde{\mathbf{c}}_i - \tilde{\mathbf{r}} \pmod{p_i}$
6:   $\tilde{\mathbf{c}}'_i \leftarrow \mathrm{Mod}\left([p^{-1}]_{p_i} \cdot \tilde{\mathbf{c}}'_i, p_i\right)$     ▷ MS Module
7: **end for**

---

Figure 3 illustrates the full architecture of NTT module. From the functionality perspective, the architecture follows Algorithm 3. At a high-level, the NTT module computes NTT of a polynomial of size $n$ in $\log n$ stages. In each stage, the module computes the transformed result of $2 \, nc_{\mathrm{NTT}}$ coefficients, thus, requiring $\frac{n}{2 \, nc_{\mathrm{NTT}}}$ steps to finish one stage.

On the three corners of the NTT architecture exist data memory, twiddle factor memories, and the output memory. At every cycle, one ME is fetched from data memory and is stored in $\mathrm{ME}_e$ and $\mathrm{ME}_o$ registers every other cycles, respectively. For each input coefficient of NTT cores, i.e., $c_{\mathrm{in.a}}^\ell$ or $c_{\mathrm{in.b}}^\ell$, a set of multiplexers select the correct coefficient from $\mathrm{ME}_e$ and $\mathrm{ME}_o$ (depicted as light blue boxes in Figure 3).

The throughput is proportional to the number of NTT cores. We denote the number of NTT cores as $nc_{\mathrm{NTT}}$. Ideally at each clock cycle, and given full utilization of NTT cores, $2 \, nc_{\mathrm{NTT}}$ coefficients are transformed. In order to read and write $2 \, nc_{\mathrm{NTT}}$ coefficients at each clock cycle, we store each polynomial across many parallel BRAMs that share common read/write address signals as depicted in Figure 3 (similar to the MULT module). This is possible thanks to the *aligned* access pattern in NTT: while access pattern changes during NTT, the number of consecutive accesses to the polynomial is always a power of two. Next, we discuss the details of the access patterns in NTT followed by our proposed solution to select each coefficient efficiently.

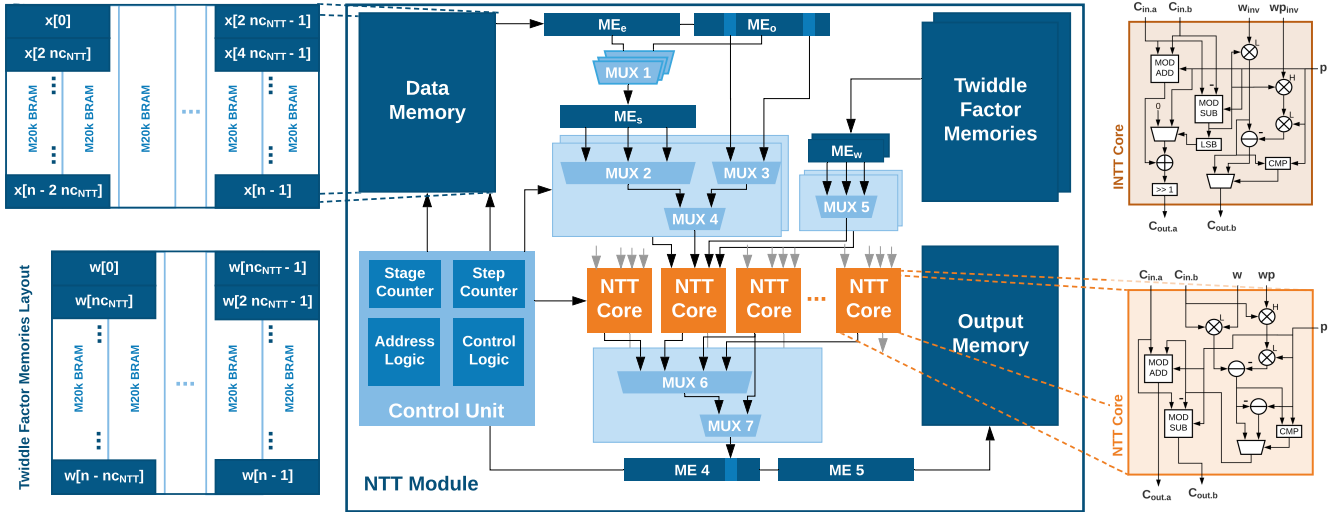

**Figure 3.** Architecture of NTT module.

### 4.3 Access Pattern

One of the main challenges in realizing the proposed NTT architecture is that the access pattern of the coefficients changes from one stage to another. We categorize the access patterns into two groups as illustrated in Figure 4. During the first (log $n - \log nc_{\text{NTT}} - 1$) stages, each pair of coefficients for each NTT core are stored in different MEs. Let us call these *Type 1* stages. For instance, consider $n = 4096$ and $nc_{\text{NTT}} = 8$, during the first step of the first stage of NTT, $x[0]$ (in $\text{ME}_0$) and $x[2048]$ (in $\text{ME}_{256}$) should be passed to the first NTT core. More precisely, polynomial coefficient $x[j]$ ($j = 0, 1, \ldots, \frac{n}{2} - 1$) is passed together with $x[j + \frac{n}{2}]$ to a given NTT core. In general, during $i^{th}$ stage, $x[j + m]$ ($j = 0, 1, \ldots, \frac{n}{2^{1+i}} - 1$) is passed along with $x[j + m + \frac{n}{2^{1+i}}]$ where $m \in \{ \frac{h \cdot n}{2^i} \mid h = 0, 1, \ldots, i \}$. The address of the ME that is fetched in Type 1 stages is computed in *Address Logic*. As soon as $\frac{n}{2^i} = 2\, nc_{\text{NTT}}$, the inter-ME data dependency no longer exists, and pairs of coefficients are selected from within each ME independently, i.e., *Type 2* stages.

In Type 1 stages, coefficients within two fetched MEs are always accessed in the same order. For example, the second coefficient in each ME is always passed to the second NTT core. However, in Type 2 stages, a coefficient at specific location of ME is passed to a different NTT core or even different inputs of an NTT core. Therefore, coefficients have to be *reordered* to be passed to NTT cores. Later in this section, we discuss an efficient method for this task.

The access pattern for twiddle factors, i.e., Y and Y′ in Algorithm 3, is different. At stage $i$, only $2^i$ *unique* values of twiddle factors, starting at index $2^i$ of twiddle polynomial, are used. Since in the worst-case scenario, $nc_{\text{NTT}}$ unique twiddle factors are used in a single *step* of NTT, we store twiddle factors in batches of size $nc_{\text{NTT}}$ in parallel.

### 4.4 Reordering Coefficients and Optimal MUXs

During Type 1 stages, once the ME is fetched, passing each coefficient within ME to the right NTT core (and right input wire) is straightforward and it can be summarized as follows:

$$\begin{cases} c_{\text{in.a}}^{\ell} = \text{ME}_e[\ell + (j \bmod 2) \cdot nc_{\text{NTT}}] \\ c_{\text{in.b}}^{\ell} = \text{ME}_o[\ell + (j \bmod 2) \cdot nc_{\text{NTT}}] \end{cases}$$

where $c_{\text{in.a}}^{\ell}$ (respectively $c_{\text{in.b}}^{\ell}$) is the input coefficient *a* (respectively *b*) of $\ell^{th}$ NTT core, $\text{ME}_e$ (resp. $\text{ME}_o$) is the memory element at "even" ("odd") read cycles, i.e., $j \bmod 2 = 0$ ($j \bmod 2 = 1$) where $j$ is the step number. In other words, $c_{\text{in.a}}^{\ell}$ (resp. $c_{\text{in.b}}^{\ell}$) is selected from one of the two positions from $\text{ME}_e$ (resp. $\text{ME}_o$) using multiplexer #3 (MUX3).

In Type 2 stages, first one of the $\text{ME}_e$ or $\text{ME}_o$ is selected using $2\, nc_{\text{NTT}}$-many two-to-one multiplexers (MUX1) and is stored in $\text{ME}_s$ registers. Next, $c_{\text{in.a}}^{\ell}$ (or $c_{\text{in.b}}^{\ell}$) receive data from one of the coefficients in $\text{ME}_s$ depending on the value of $\ell$ and $i$. The naive approach is to use one multiplexer per each coefficient input of every NTT core that selects one number from $2\, nc_{\text{NTT}}$ fetched numbers. We denote such multiplexer as $\text{MUX}^{2\, nc_{\text{NTT}}}$. As a result, we need $2\, nc_{\text{NTT}}$-many $\text{MUX}^{2\, nc_{\text{NTT}}}$ to pass coefficients to NTT cores and the same number of MUXs to reorder them to be written back to the memory.

These MUXs not only make the placement and route process more challenging but also consume enormous number of registers and logic blocks. Moreover, scaling the NTT module to higher number of cores (> 32) is inefficient due to super-linear resource consumption with respect to $nc_{\text{NTT}}$. In our case, synthesis tools failed to place and route the required resources to realize these MUXs. In contrast, we take advantage of the observation that *NTT cores' inputs have a different number of possibilities from which they select the correct coefficient at a given stage*. For example, during Type

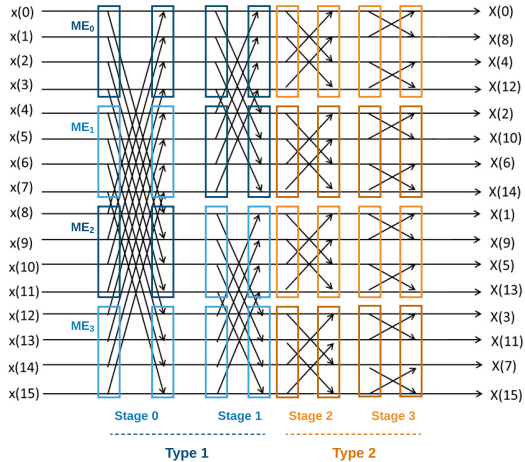

**Figure 4.** Access pattern of Type 1 and Type 2 stages in NTT. 2 stages, $c_{in.a}^0$ only receives coefficients from the first word of the fetched ME, regardless of the stage or step number.

In the worst-case scenario, there are $\log nc_{NTT}$ different indices from which a coefficient should be accessed from $ME_s$ for a particular NTT core input. Therefore, instead of using $(4 \cdot nc_{NTT})$-many $\text{MUX}^{2\,nc_{NTT}}$, we instantiate $(4 \cdot nc_{NTT})$-many MUXs of size at most $\text{MUX}^{\log 2\,nc_{NTT}}$. These optimal multiplexers are shows as MUX2 in Figure 3. The selection signal of these MUXs is set to $s = \log n - 1 - i$ ($i$ being the stage number). The corresponding inputs ($\text{MUX}\{c_{in.a}^\ell\}(\alpha)$ and $\text{MUX}\{c_{in.b}^\ell\}(\alpha)$) from which a coefficient should be selected are assigned based on the following formula:

$$= \begin{cases} ME_s[(\ell \,\&\, (2^s - 1)) + ((i >> s) << (s + \ell))] \\ ME_s[(\ell \,\&\, (2^s - 1)) + ((i >> s) << (s + \ell)) + 2^s] \end{cases}$$

where $\text{MUX}\{c_{in.a}^\ell\}(\alpha)$ is the $\alpha$-th input wire of the MUX that selects the corresponding input coefficient from $ME_s$ for the $\ell$-th core, thus, $0 \leq \alpha < \log nc_{NTT}$. Finally, depending on the stage type, MUX4 selects the output of MUX2 or MUX3.

A similar set of MUXs (MUX6 and MUX7) are used to reorder the data back before storage. Final results (ME4 and ME5) will be stored in the data memory during two consecutive clock cycles; except for the the last stage where they will be stored in *output memory*. The optimized multiplexers for twiddle factors is designed in a similar manner. The optimal multiplexers are an integral part of the design of NTT module. For instance, this optimization reduces the number of registers used for a 8-core NTT module from 224,000 to 97,000. Note that to synthesize the NTT module, register is the most limited resource (see Section 7), thus, without the proposed optimal multiplexers, one cannot scale the design properly.

### 4.5 NTT High-Level Pipeline

Storing polynomial coefficients across parallel memory blocks enables simultaneous access to multiple coefficients. However, the NTT cores cannot be fully utilized due to the following reason. During Type 1 stages, coefficients that should be passed to each NTT core are not located in the same ME. Therefore, two different MEs should be read before the computation can start which introduces 50% bubble in the NTT core pipeline. More precisely, first $\log n - \log nc_{NTT} - 1$ stages face this problem. Given that NTT modules consume most of the FPGA resources, this issue reduces the throughput of the entire design to $(\log n - \log nc_{NTT} - 1)/\log n$.

To address this problem, we propose to double the size of MEs and store $2\,nc_{NTT}$ consecutive coefficients in each memory element. Meanwhile, we reduce the depth of the memories that store the polynomial by half. Even though it is still necessary to read two MEs before starting the computation, we can now transform *two* MEs in the next two cycles and store them back in the memory. This modification results in the full utilization of NTT core. In order to have minimal BRAM usage, all of the reads and writes during different NTT stages are *inplace*, and no additional BRAM is used to store intermediate values.

### 4.6 Memory Utilization and Word-Packing

Storing multiple polynomial coefficients in multiple parallel memory units (M20K) causes memory block underutilization both depth-size and width-size. Consider a general scenario where $\beta$-many numbers are stored in parallel:

*Depth-wise:* Each M20K memory unit holds 512-many 40-bit wide words and at any cycle, one word can be read from or written into the memory. When fewer than 512 words are stored in the memory, the rest of the memory rows cannot be used to store a secondary polynomial since at any point in time we are reading/writing one word associated with the first polynomial. As long as $\frac{n}{\beta} \geq 512$, M20K is fully utilized. This inequality generally holds in our architecture except when $n = 2^{12}$ (smallest polynomial size) and $nc_{NTT} = 16$ which makes M20K *half* utilized. However, this is not an issue since our design is not BRAM-constrained when $n = 2^{12}$.

*Width-wise:* As the polynomial-size ($n$) grows, our design becomes more and more BRAM-constrained to the extent that at $n = 2^{14}$, there is not enough BRAM on the chip; thus, we have to use DRAM as well (we will discuss this in more detail in Section 6). Therefore, it is essential that the polynomials are efficiently stored in memory. By storing each coefficient in a separate physical BRAM, we will only reach $\frac{54}{2 \cdot 40} = 68\%$ utilization. In contrast, we pack multiple coefficients and store them in fewer M20K units as shown in Figure 3 reaching memory utilization of $\beta \cdot 54/(\lceil \beta \cdot 54/40 \rceil \cdot 40)$. For $\beta = 8$, BRAM utilization will reach more than 98%.

**Performance.** Computing the NTT of a polynomial requires $\log n$ stages and each stage takes $\frac{n}{2\,nc_{NTT}}$ cycles. Hence, it takes $\frac{n \log n}{2\,nc_{NTT}}$ cycles to compute one NTT.

**INTT Module.** This module is identical to the NTT module except: (i) the NTT core is replaced by the INTT core, (ii) the control unit operates in the reverse order of stage numbers, and (iii) twiddle factors correspond to the INTT calculations.

# 5 KEYSWITCH MODULE

In this section, we discuss the KeySwitch algorithm followed by our proposed architecture and the design details.

## 5.1 Algorithm

Key switching is a technique to make a ciphertext decryptable with a different secret key homomorphically. Various gadget decomposition methods can be adopted to balance noise growth and execution time. Given $q_{d-1}$, the product of coprime integers $p_0, \ldots, p_{d-1}$, and $q_\ell$ divides $q_{d-1}$, define gadget decomposition $R_{q_\ell} \mapsto R^d$ as $\mathbf{g}^{-1}(\mathbf{a}) = \left([\mathbf{a}]_{p_i}\right)_{0 \le i \le d-1}$, and gadget vector as $\mathbf{g} = \left(\pi_i \left[\pi_i^{-1}\right]_{p_i}\right)_{0 \le i \le d-1}$ where $\pi_i = \frac{q_{d-1}}{p_i}$. This choice of gadget decomposition contributes to a fast key switching and high noise growth. With the special modulus $p$ and a rescaling at the end of key switching, explained in [15], key switching is almost noise-free.

- KeySwitch(ct, ksk): Given a ciphertext $\text{ct} = (\mathbf{c}_0, \mathbf{c}_1) \in R_{q_\ell}^2$ decryptable with secret key $\mathbf{s}$ and a key switching key $\text{ksk} = (\mathbf{D}_0 \mid \mathbf{D}_1) \in R_{q_\ell p}^{(L+2) \times 2}$, where | appends one column vector to another, generate a new ciphertext $\text{ct}' = (\mathbf{c}_0', \mathbf{c}_1') \in R_{q_\ell}^2$ decryptable with secret key $\mathbf{s}'$. (see Algorithm 5).

## 5.2 KeySwitch Architecture

KeySwitch is the most computationally intensive *high-level* operation in CKKS. It has several important roles, including relinearization and ciphertext rotation. Figure 5 illustrates the KeySwitch architecture, which from the functionality perspective corresponds to Algorithm 5. To reduce on-chip memory usage, our design takes one polynomial (one RNS component) at a time and outputs two polynomials. Recall that in CKKS, all polynomials are in NTT form by default. Thus, once the input polynomial is written into the input memory, it has to be converted back to the original domain. This process is performed using the first INTT module (INTT0). Next, the polynomial is transformed to the NTT form for all other primes (including the special modulus).

Since per each INTT computation, we have to perform $k$ NTT, the throughput of the NTT module(s) has to be $k$-times the throughput of INTT0. Here, $k$ is the number of RNS components of ciphertext modulus, i.e., $L + 1$. This requirement can be realized in two different ways: (i) having one NTT module with $k$-many more cores than INTT0 or (ii) having multiple NTT module with fewer cores per each module. We denote this NTT module (or a set of them) as NTT0. We will discuss the trade-offs later in this section. In Figure 5, the second approach (using more than one NTT module) is chosen for $n = 2^{13}$ and $k = 4$ parameter set.

Once the NTT computations are finished, the DyadMult module computes the dyadic product between the output of NTT modules and the relinearization/Galois keys according to Algorithm 5. Recall that a dyadic product on the original input polynomial is also needed in KeySwitch; thus, a

---

**Algorithm 5** Key Switching | KeySwitch(ct, ksk)

**Input:** $\text{ct} = (\tilde{\mathbf{C}}_0, \tilde{\mathbf{C}}_1) \in (\prod_{i=0}^{\ell} R_{p_i})^2$, and $\text{ksk} = \left(\left(\tilde{\mathbf{D}}_{i,0}\right)_{0 \le i \le L+1} \mid \left(\tilde{\mathbf{D}}_{i,1}\right)_{0 \le i \le L+1}\right) \in (p \prod_{i=0}^{L} R_{p_i})^{(L+2) \times 2}$

**Output:** $\text{ct}' = (\tilde{\mathbf{C}}_0', \tilde{\mathbf{C}}_1') \in (\prod_{i=0}^{\ell} R_{p_i})^2$

1: **for** ($i = 0$; $i \le \ell$; $i = i + 1$) **do**
2:   $\bar{\mathbf{a}} \leftarrow \text{INTT}_{p_i}(\tilde{\mathbf{c}}_{1,i})$    ▷ INTT Module
3:   **for** ($j = 0$; $j \le \ell$; $j = j + 1$) **do**
4:     **if** $i \ne j$ **then**
5:       $\bar{\mathbf{b}} \leftarrow \text{Mod}(\bar{\mathbf{a}}, p_j)$
6:       $\tilde{\mathbf{b}} \leftarrow \text{NTT}_{p_j}(\bar{\mathbf{b}})$    ▷ NTT Module
7:     **else**
8:       $\tilde{\mathbf{b}} \leftarrow \tilde{\mathbf{a}}$
9:     **end if**
10:     $\tilde{\mathbf{c}}_{0,j}'' \leftarrow \tilde{\mathbf{c}}_{0,j}'' + \tilde{\mathbf{b}} \odot \tilde{\mathbf{d}}_{i,0,j}$ (mod $p_j$)
11:     $\tilde{\mathbf{c}}_{1,j}'' \leftarrow \tilde{\mathbf{c}}_{1,j}'' + \tilde{\mathbf{b}} \odot \tilde{\mathbf{d}}_{i,1,j}$ (mod $p_j$) ▷ Dyadic Mod.
12:   **end for**
13:   $\bar{\mathbf{b}} \leftarrow \text{Mod}(\bar{\mathbf{a}}, p)$
14:   $\tilde{\mathbf{b}} \leftarrow \text{NTT}_p(\bar{\mathbf{b}})$    ▷ NTT Module
15:   $\tilde{\mathbf{c}}_{0,\ell+1}'' \leftarrow \tilde{\mathbf{c}}_{0,\ell+1}'' + \tilde{\mathbf{b}} \odot \tilde{\mathbf{d}}_{0,i,L+1}$ (mod $p_j$)
16:   $\tilde{\mathbf{c}}_{1,\ell+1}'' \leftarrow \tilde{\mathbf{c}}_{1,\ell+1}'' + \tilde{\mathbf{b}} \odot \tilde{\mathbf{d}}_{1,i,L+1}$ (mod $p_j$) ▷ Dyd. M.
17: **end for**
18: $\text{ct}' \leftarrow (\text{Floor}(\tilde{\mathbf{C}}_0'', p), \text{Floor}(\tilde{\mathbf{C}}_1'', p))$ ▷ INTT/NTT/MS
19: $\text{ct}' \leftarrow \text{CKKS.Add}(\text{ct}, \text{ct}')$

---

separate Dyadic module is used. After dyadic product computation, the result is stored in the corresponding memory banks. There are two *sets* of BRAM banks, each bank containing the RNS components of one polynomial.

The computation flow described above repeats for $k$-many times (one per each RNS component). The result is accumulated in the BRAM banks. After $k$ iterations, the second part of the computation – usually referred to as *Modulus Switching* (developed in [12]) – is performed. In Modulus Switching which executes Floor, the polynomial corresponding to the *special modulus* has to be transformed back to the time domain (by INTT1) and then be transformed using every other $k$ primes (by NTT1). The aforementioned process is independently performed for both sets of banks. Next, the polynomial is multiplied by the inverse value of the associated prime and subtracted from the result of the first half of KeySwitch computation. The MS module embeds multiplication and subtraction operations. The output of KeySwitch is stored as two sets of $k$ polynomials referred to as "Output Poly 0/1".

## 5.3 Balancing Throughput

Our primary goal in designing KeySwitch architecture is to have a fully *end-to-end pipelined* module that can process many key switching operations simultaneously without excessive FIFOs between different components. Thus, we have to tune the throughput of each component carefully. As we discussed in Section 4, this is one of the reasons to design a flexible architecture for NTT, the throughput of which can be adjusted. According to Algorithm 5, per each initial INTT, we have to compute $k$ NTTs. In general, let's denote the number of NTT0 as $m_0$ (assuming a power of two number). Thus, we have: $nc_{\text{NTT}_0} = k \cdot nc_{\text{INTT}_0}/m_0$.

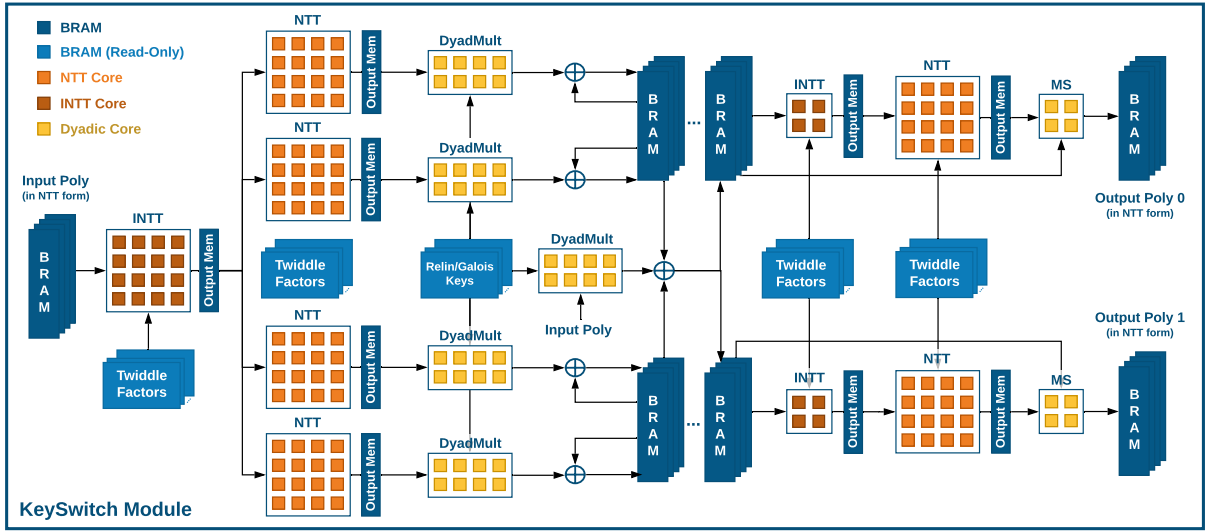

**Figure 5.** Architecture of KeySwitch module.

Next, we compute the number of cores needed for DyadMult. Recall that it takes $(n \log n)/(2\, nc_{\text{NTT}_0})$ cycles for NTT module to finish the computation. The DyadMult module has to compute the product of NTT output with two different sets of keys (ksk = $D_0 \mid D_1$). It takes $(2\,n)/nc_{\text{DYD}}$ cycles to perform dyadic multiplication on the output of the NTT module. Since in general, $\log n$ is not a power of two, the throughputs do not perfectly match. We make sure that the throughput of Dyadic module is greater than that of (or equal to) the NTT module by satisfying the following inequality:

$$\frac{2\,n}{nc_{\text{DYD}}} \leqslant \frac{n \log n}{2\, nc_{\text{NTT}_0}} \Rightarrow nc_{\text{DYD}} = \left\lceil \frac{4\, nc_{\text{NTT}_0}}{\log n} \right\rceil$$

The throughput of INTT1 modules can be adjusted by assigning $nc_{\text{INTT}_1} = \lceil nc_{\text{INTT}_0}/k \rceil$. One can also determine $nc_{\text{NTT}_1} = nc_{\text{INTT}_0}$ and $nc_{\text{MS}} = \lceil (2\, nc_{\text{NTT}_1})/\log n \rceil$. For two FPGA chips that we have implemented HEAX on, the optimal architecture parameters are computed and summarized in Table 5.

### 5.4 KeySwitch Ops. and Synchronization

Figure 6 shows the high-level pipeline of KeySwitch module for $n = 2^{13}$ (third row of Table 5). All of the modules – and their internal components – are pipelined, and the throughput is balanced. As can be seen, multiple key switching operations are computed simultaneously in different pipeline stages (in lighter colors). The fifth Dyadic module that operates on input polynomial BRAM is *synchronized* with the rest of the Dyadic modules even though the computation can be started as soon as the input poly is available. The reason is that during each of the $k$ iterations of Dyadic product, each module computes and accumulates the results by reading/writing from/to a separate BRAM bank. This enables us to avoid any *memory replication* considering that these memory banks are prohibitively large. However, this *delayed* computation introduces a dependency problem in the pipeline referred to as "Data Dependency 1". By the time the $k$-th Dyadic module starts the computation, the content of input poly is overridden by the next key switching operation. As a result, we allocate enough BRAM to hold $f_1$-many polynomials where $f_1 = \left\lceil 3 + \frac{nc_{\text{INTT}_0}}{nc_{\text{NTT}_0}} \right\rceil$. Similarly, MS module receives inputs from DyadMult modules. This is marked as "Data Dependency 2" in Figure 6. Thus, we need to allocate more memory to store the output of the DyadMult modules in $f_2$ different buffers. The value of $f_2$ can be computed as: $f_2 = \left\lceil 1 + m_0 \cdot \frac{nc_{\text{INTT}_1}}{nc_{\text{NTT}_1}} + \frac{nc_{\text{INTT}_1} \cdot \log n}{nc_{\text{MS}}} \right\rceil$.

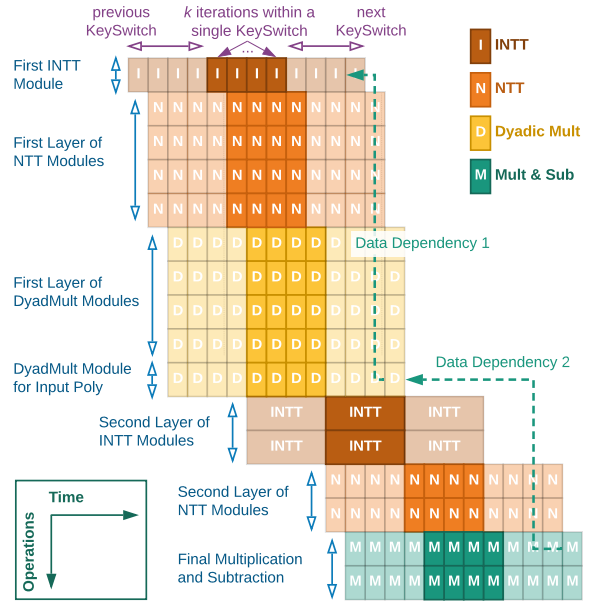

**Figure 6.** High-level pipeline of KeySwitch module.

# 6 SYSTEM-VIEW and DATA FLOW

In this section, we discuss a higher-level view of the computation and elaborate on the data flow. Figure 7 shows a system-view comprising host CPU and FPGA Board which are connected via Peripheral Component Interconnect express (PCIe) bus. On FPGA board, exist the FPGA chip as well as off-chip DRAM memory connected via DDR interface.

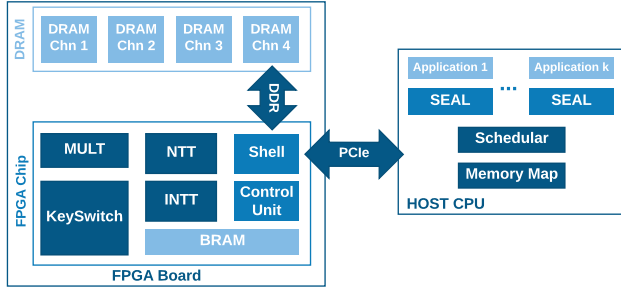

**Figure 7.** System-view of HEAX.

## 6.1 On-Chip vs. Off-Chip Memory Accesses

There are two main ways to store data on FPGA board: (i) *off-chip* DRAM with several Gigabytes of capacity but high response delay and (ii) *on-chip* BRAM with few megabits of capacity but very fast response time and high throughput. As has been shown by prior art [53, 54], leveraging off-chip memory to store intermediate results significantly reduces the overall performance due to high delays between subsequent reads and writes. One of our primary design goals is to avoid off-chip memory access as much as possible. We have introduced several techniques to use minimal on-chip memory, re-use many BRAM units, together with data compaction (see Section 4 and Section 5). As a result, no off-chip memory access is performed for $n = 2^{12}$ parameter set on both Arria 10 and Stratix 10 FPGAs; which is one of the main reasons for our unprecedented performance improvements.

For $n = 2^{13}$ parameter set, there is sufficient on-chip memory on Stratix 10 chip. Unfortunately, for $n = 2^{14}$, there is not enough BRAM available for our design, and as a result, we have to move some part of the data to off-chip memory. In order to minimize the effect of off-chip memory accesses, we choose to put key switching keys (ksk) in DRAM because of two main reasons: (i) the size of these keys grow very rapidly with HE parameters. In general, the size of ksk grows as $O(n k^2)$, and roughly speaking, $k$ grows linearly with $n$ which results in (almost) $O(n^3)$. This is the highest growth rate compared to all other memory components, including twiddle factors which grow as $O(n k)$. (ii) ksk is only read once per each KeySwitch. Note that each unique element of twiddle factors is read $k$ times during one KeySwitch operation; thus, twiddle factors are less suitable candidates.

We distribute the ksk among four different DRAM banks such that at any point in time, the full capacity of off-chip memory bandwidth is used. In order to further mask the effect of DRAM accesses, we leverage the *burst mode* in which a long sequence of data is read at the same time on each channel. The entire process of reading ksk from DRAM is pipelined to minimize the drop in throughput of KeySwitch. It is worth-mentioning that DRAM bandwidth is sufficient to match the throughput of KeySwitch. Per each KeySwitch, two sets of ksk have to be streamed to FPGA chip. Each of these sets, hold $k \cdot (k + 1)$-many vectors of size $n$. Substituting $n = 2^{14}$, $k = 8$, and 64-bit per each word results in approximately 151 megabits. We have to stream this volume of data in 383 microseconds (please see Table 8). Therefore, DRAM bandwidth should be higher than 49.28 GBps, which is indeed lower than the measured bandwidth of all four channels combined.

In addition to storing ksk, we use DRAM for one more purpose. In some applications, it is more efficient to store the result of computation in DRAM instead of sending them back to CPU (in case these results are going to be used soon). The address at which the result is stored is held on the CPU side and is shown as "Memory Map". The memory map is used to point to the ciphertext(s) that are stored in DRAM to be used later on without involving PCIe.

## 6.2 Data Transfer on PCIe

In order to maximize the utilization of computation blocks on FPGA, we need to interleave computation and data transfer between FPGA and CPU. We divide this design process into two parts: CPU-side and FPGA-side. On the CPU-side, we need to sequence and batch multiple operations in the program (that uses SEAL) and start the data transfer process on PCIe using *multiple threads*. On the FPGA-side, we need to allocate the necessary buffers to store the received data. In what follows, we explain these two parts in more detail.

**Sequencing and Batching.** Transferring data on PCIe involves three main steps: (i) a memcpy is issued to copy the content of the polynomial to pinned memory pages, (ii) CPU signals FPGA that the data is ready, and (iii) FPGA reads the data from PCIe. In order to reduce the data copy time, Direct Memory Access (DMA) is used. However, even by relying on DMA, the maximum throughput that PCIe can provide depends on the message size and the number of simultaneous data transfer requests. Therefore, we transfer (at least) a complete polynomial ($2^{15} - 2^{17}$ Bytes) in each request. Moreover, we implement a multi-threaded data transfer mechanism that uses eight threads to interleave eight separate polynomials at a time to maximize the PCIe throughput and avoid unnecessary bubbles in the computation pipeline.

**Double and Quadruple Buffering.** For the MULT module, it suffices to double-buffer the input such that CPU writes to one of these buffers and FPGA reads from the other one. For KeySwitch module, however, we need to perform quadruple buffering due to the data dependency on input polynomial as discussed in Section 5. In order to make sure buffers are not overridden before they are read, we stop the writing process if the buffer has not been read yet.

## 7 IMPLEMENTATION and EXPERIMENTS

### 7.1 Experimental Setup

In this section, we discuss the resource consumption of HEAX components as well as the performance comparison with CPUs and GPUs. To illustrate the adaptability of HEAX, we implement HEAX on two FPGAs which represent two different classes of computational resources. Table 1 summarizes the breakdown of resources of each FPGA chip. There are three major types of resources that are available:

• Digital Signal Processing (DSP) units that are able to perform one 27-bit or two 18-bit multiplications.
• Adaptive Logic Modules (ALM) are core logic units with two combinational adaptive look-up tables, a two-bit full adder, and four 1-bit Registers (REG).
• Block RAM (BRAM) units that are on-chip memories. Each M20K unit of BRAM holds 512-many 40-bit values.

| Board | Chip | Chip Resources | | | | | DRAM | |
|---|---|---|---|---|---|---|---|---|
| | | DSP | REG | ALM | BRAM bits | BRAM #M20K | #chnl. | BW (GBps) |
| Board-A | Arria 10 GX 1150 | 1518 | 1.71M | 427K | 53Mb | 2.7K | 2 | 34 |
| Board-B | Stratix 10 GX 2800 | 5760 | 3.73M | 933K | 229Mb | 11.7K | 4 | 64 |

**Table 1.** Summary of FPGA boards' specifications.

### 7.2 FHE Parameters and Security Guarantees

The security guarantees of HEAX directly derives from the CKKS scheme [17] since the functionality of the scheme is not altered. The security parameters for which we have instantiated HEAX are borrowed from the HE security standards [2] for 128-bit classical security. Changing the underlying word-size in HEAX reduces the number of DSPs used but does not affect the security since the total bitwidth of the ciphertext modulus is preserved [2]. Similarly, we leveraged the RNS-level parallelism which is proven to be secure [16].

We evaluate our design on a wide range of FHE parameters: from ciphertext polynomial size ($n$) of $2^{12}$ and 109-bit ciphertext modulus ($\lfloor \log qp \rfloor + 1$) to $2^{14}$ with 438-bit ciphertext modulus. We refer to these parameter sets as Set-A, Set-B, and Set-C, respectively (summarized in Table 2). Recall that $k$ is the number of small RNS components of ciphertext modulus. Parameters with 128-bit post-quantum security require slightly smaller ciphertext moduli. We select as few prime moduli for RNS as possible for superior performance [36]. Note that parameter sets corresponding to $2^{11}$ (or lower) are almost never used in practice due to the multiplication depth of 1 (or zero). Choosing $2^{15}$ (or higher) results in enormous computation blow-up and are also rarely used in practice.

| HE Param. Set | $n$ | $\lfloor \log qp \rfloor + 1$ | $k$ |
|---|---|---|---|
| Set-A | $2^{12}$ | 109 | 2 |
| Set-B | $2^{13}$ | 218 | 4 |
| Set-C | $2^{14}$ | 438 | 8 |

**Table 2.** The HE parameter sets used in this paper. $n$ is the ciphertext polynomial size, $qp$ is the ciphertext modulus, and $k$ is the number of RNS components of $q$.

### 7.3 Resource Consumption

**Computation Cores.** Table 3 provides a detailed resource consumption of Dyadic, NTT, and INTT computation cores as well as the number of pipeline stages (delay) for each core.

| Core Name | DSP | REG | ALM | #Stages |
|---|---|---|---|---|
| Dyadic | 22 | 4526 | 1663 | 23 |
| NTT | 10 | 6297 | 2066 | 50 |
| INTT | 10 | 5449 | 2119 | 49 |

**Table 3.** Resource consumption of each computation core.

**Basic Modules.** Table 4 provides a detailed resource consumption of different modules (with various number of cores). The BRAM utilization is reported for Set-B parameters ($n = 2^{13}$). The BRAM *bits* usage in each module does not depend on the number of cores but the number of M20K units does. The reason is that more coefficients are stored in parallel M20K units. In the last column, the number of cycles that takes for each module to process a polynomial (or pair of polynomials in case of MULT module) is reported.

| Module | #Cores | DSP | REG | ALM | BRAM #bits | BRAM #M20K | Cycles |
|---|---|---|---|---|---|---|---|
| A10 Shell | - | 1 | 79203 | 39222 | 886496 | 144 | - |
| S10 Shell | - | 2 | 86984 | 45612 | 1201096 | 173 | - |
| MULT | 4 | 88 | 42817 | 15795 | 1104384 | 65 | 1024 |
| | 8 | 176 | 61878 | 22160 | | 65 | 512 |
| | 16 | 352 | 93594 | 35257 | | 164 | 128 |
| | 32 | 704 | 181503 | 62157 | | 293 | 64 |
| NTT | 4 | 40 | 61670 | 22316 | 1514496 | 86 | 6144 |
| | 8 | 80 | 96919 | 36336 | | 185 | 3072 |
| | 16 | 160 | 196205 | 67865 | | 380 | 1536 |
| | 32 | 320 | 387357 | 142300 | | 725 | 768 |
| INTT | 4 | 40 | 63917 | 22700 | 1514496 | 86 | 6144 |
| | 8 | 80 | 104575 | 37331 | | 185 | 3072 |
| | 16 | 160 | 182478 | 68645 | | 380 | 1536 |
| | 32 | 320 | 384267 | 144957 | | 724 | 768 |

**Table 4.** Resource consumption of basic modules.

**Complete Design.** Table 6 provides a breakdown of FPGA resource consumption for different HE parameter sets. The complete design encompasses the KeySwitch module along with the MULT module. For standalone NTT requests from CPU, the NTT modules within KeySwitch is used.

### 7.4 Performance

**Critical Paths and Maximum Clock Frequency.** We have analyzed the critical paths of our design and have eliminated such paths during many design iterations reaching the maximum clock frequency of 275 MHz and 300 MHz for Arria 10 and Stratix 10 FPGA chips, respectively.

**Scalability.** One of design principles of HEAX is that it can automatically be instantiated at different scales with no manual tuning, enabling cloud providers to seamlessly use HEAX based on the underlying hardware resource. To illustrate this, we have instantiated HEAX for the *same HE parameters* (Set-A) but at two *different scales* (see Table 5). The up-scaled version on Stratix 10 consumes (close to) twice the resources (Table 6) and provides twice the throughput compared to Arria 10 instantiation (see Table 8).

| FPGA Device | HE Param. Set | KeySwitch Architecture Parameter Set |
|---|---|---|
| Arria10 | $n = 2^{12}$ (Set-A) | $1 \times \text{INTT}^{(8)} \to 2 \times \text{NTT}^{(8)} \to 3 \times \text{Dyad}^{(4)} \to 2 \times \text{INTT}^{(4)} \to 2 \times \text{NTT}^{(8)} \to 2 \times \text{MS}^{(2)}$ |
| | $n = 2^{12}$ (Set-A) | $1 \times \text{INTT}^{(16)} \to 2 \times \text{NTT}^{(16)} \to 3 \times \text{Dyad}^{(8)} \to 2 \times \text{INTT}^{(8)} \to 2 \times \text{NTT}^{(16)} \to 2 \times \text{MS}^{(4)}$ |
| Stratix10 | $n = 2^{13}$ (Set-B) | $1 \times \text{INTT}^{(16)} \to 4 \times \text{NTT}^{(16)} \to 5 \times \text{Dyad}^{(8)} \to 2 \times \text{INTT}^{(4)} \to 2 \times \text{NTT}^{(16)} \to 2 \times \text{MS}^{(4)}$ |
| | $n = 2^{14}$ (Set-C) | $1 \times \text{INTT}^{(8)} \to 4 \times \text{NTT}^{(16)} \to 5 \times \text{Dyad}^{(8)} \to 2 \times \text{INTT}^{(1)} \to 2 \times \text{NTT}^{(8)} \to 2 \times \text{MS}^{(4)}$ |

Table 5. KeySwitch architecture for different HE parameter sets.

| FPGA Device | HE Param. Set | DSP (%) | REG (%) | ALM (%) | BRAM bits (%) | BRAM #M20K (%) | Freq. (MHz) |
|---|---|---|---|---|---|---|---|
| Arria10 | Set-A | 1185 (78) | 723188 (42) | 246323 (58) | 26596320 (48) | 1731 (64) | 275 |
| | Set-A | 2018 (35) | 1554005 (42) | 582148 (62) | 26907592 (11) | 3986 (34) | 300 |
| Stratix10 | Set-B | 2610 (45) | 1976162 (53) | 698884 (75) | 201332624 (84) | 10340 (88) | 300 |
| | Set-C | 2370 (41) | 1746384 (47) | 599715 (64) | 182847524 (76) | 9329 (80) | 300 |

Table 6. Resource consumption of HEAX for different HE parameter sets.

| FPGA Device | HE Param. Set | NTT | | | INTT | | | Dyadic MULT | | |
|---|---|---|---|---|---|---|---|---|---|---|
| | | CPU | HEAX | Speed-up | CPU | HEAX | Speed-up | CPU | HEAX | Speed-up |
| Arria10 | Set-A | 7222 | 89518 | 12.4 | 7568 | 89518 | 11.8 | 36931 | 1074219 | 29.1 |
| | Set-A | 7222 | 195313 | 27.0 | 7568 | 195313 | 25.8 | 36931 | 1171875 | 31.7 |
| Stratix10 | Set-B | 3437 | 90144 | 26.2 | 3539 | 90144 | 25.5 | 18362 | 585938 | 31.9 |
| | Set-C | 1631 | 41853 | 25.7 | 1659 | 41853 | 25.2 | 9117 | 292969 | 32.1 |

Table 7. Performance comparison of HEAX with CPU. Number of operations per second for CKKS *low-level* operations.

| FPGA Device | HE Param. Set | KeySwitch | | | MULT+ReLin | | |
|---|---|---|---|---|---|---|---|
| | | CPU | HEAX | Speed-up | CPU | HEAX | Speed-up |
| Arria10 | Set-A | 488 | 44759 | 91.7 | 420 | 44759 | 106.6 |
| | Set-A | 488 | 97656 | **200.5** | 420 | 97656 | **232.5** |
| Stratix10 | Set-B | 97 | 22536 | **232.3** | 84 | 22536 | **268.3** |
| | Set-C | 16 | 2616 | **163.5** | 15 | 2616 | **174.4** |

Table 8. Performance comparison of HEAX with CPU. Number of operations per second for CKKS *high-level* operations.

**Performance Comparison with GPUs.** To the best of our knowledge, there does not exist any work based on FPGAs or GPUs for CKKS scheme. In Table 9, we compare the performance of our "NTT architecture" on Stratix10 (which holds ten 16-core NTT modules) with two NVIDIA GPUs [1]. Not only HEAX consumes significanlty less power but it is **36–81×** faster compared to data-center GPUs.

| Polynomial Size | HEAX (10×16-cores) | Tesla-K80 (2496-cores) | Tesla-P100 (3584-cores) | Performance Comparison |
|---|---|---|---|---|
| $2^{12}$ | 1953130 | 25641 | 27777 | **70–76×** |
| $2^{13}$ | 901440 | 20833 | 25000 | **36–43×** |
| $2^{14}$ | 418530 | 5181 | 11494 | **36–81×** |

Table 9. Performance comparison (operations per second) of HEAX with NVIDIA GPUs for NTT computation.

**Performance Comparison with CPUs.** We compare the performance of HEAX with Microsoft SEAL V3.3 [56], which is an FHE library for BFV and CKKS schemes that has undergone several years of performance optimizations. We measure the performance of SEAL on a single-threaded Intel Xeon(R) Silver 4108 running at 1.80 GHz; which is a similar CPU used in prior art [54]. The single-thread baseline is used by prior art for measuring the performance (non-CKKS schemes) [54]. In addition, SEAL is thread-safe but *not multithreaded* due to the complex data dependencies, hence, we cannot compare to a multi-threaded execution. In general, CKKS evaluation functions do not have a balanced parallelizable computation flow and many parts are not parallelizable at all. For instance, the "Modulus Switching" is not parallelizable leading to the Data-Dependency 2 (Figure 6). This is the reason why we cannot allocate a single NTT/INTT module in KeySwitch and use it over time for different steps. Instead, we design an end-to-end *pipelined* design and use the chip-area proportional to the computation overhead.

Table 7 shows the performance results (number of operations per second) of HEAX for low-level operations and its comparison with SEAL. Results are reported for processing a single polynomial (in case of NTT/INTT) or pair of polynomials (MULT). On Stratix 10, 16-core modules are instantiated. On Arria 10, a 16-core MULT and 8-core NTT/INTT modules are used (see Table 5). Note that we report the performance results for low-level operation merely for completeness. These operations are rarely used in isolation and are instead used as part of high-level operations. For high-level operations, i.e., Rotation and Relinearization (using KeySwitch) and a complete ciphertext multiplication (using MULT and KeySwitch), the performance improvements are more pronounced as shown in Table 8. As can be seen, HEAX achieves close to *two orders of magnitude* performance improvement using Arria 10 compared to CPU (first row of Table 8). On a more powerful FPGA, i.e., Intel Stratix 10 (Board-B), HEAX achieves **164–268×** performance improvements among various HE parameter sets (second to fourth rows of Table 8).

## 8 RELATED WORK

The CKKS scheme is one of the most recently proposed FHE schemes that allows homomorphic operations on fixed-point numbers; making it the prime candidate for machine learning applications. To the best of our knowledge, no hardware architecture has been proposed for the CKKS scheme, and in this paper, we propose the first of its kind. As a result, it is not fair to compare the performance of HEAX with previous designs that focus on non-CKKS schemes. In what follows, we briefly review the research effort related to FPGA, ASIC, and GPU-based acceleration for non-CKKS schemes.

**Hardware Acceleration for non-CKKS Schemes.** In [53], a system based on FPGA is proposed for BFV scheme to process ciphertext polynomial sizes of $2^{15}$. However, due to the massive off-chip data transfer, their design does not yield superior performance compared to CPU execution.

Perhaps, the closest work to ours is by Roy et al. [54] in which authors propose an architecture for BFV scheme and implement their design on Xilinx Zynq UltraScale+ MPSoC ZCU102. In order to avoid off-chip memory accesses, authors focus on $n = 2^{12}$ ciphertext sizes and report 13× speed-up (using two instances of their proposed processors) compared to the FV-NFLlib [32] executing on an Intel i5 processor running at 1.8 GHz. However, compared to a more optimized Microsoft SEAL library [55], FV-NFLlib is 1.2× slower [7]. In addition, our design is significantly more modular and scalable. We have instantiated HEAX for three different set of HE parameters with no manual tuning (polynomial sizes of $2^{12}$, $2^{13}$, and $2^{14}$). Moreover, HEAX has a multi-layer pipelined design and is drastically more efficient, offering more than two orders of magnitude performance improvement compared to Microsoft SEAL running on Intel Xeon Silver 4108 at 1.8 GHz (note that similar processor is used compared with [54] running at identical frequency).

**FPGA-based Co-Processors.** Designing co-processors has also been studied in the literature. These co-processors work in conjunction with CPUs and accelerate one or more of the homomorphic operations [20, 37, 39, 41, 46, 47]. In [46] and [37, 47], authors focus on designing hardware architecture for the *encryption* operation only, by leveraging Karatsuba and Comba multiplication algorithms, respectively. In [20], a Homomorphic Encryption Processing Unit (HEPU) is proposed for LTV scheme [45]. Authors focus on accelerating the Chinese Remainder Transform (CRT) for power-of-2 cyclotomic rings and report 3.2–4.4× performance improvements for homomorphic multiplication using Xilinx Virtex-7.

**Large-Integer Multiplication Hardware Acceleration.** A line of research focuses on designing very large integer multipliers (768K-bit to 1.18M-bit multiplications) – based on FPGAs or ASICs – that can be used to accelerate homomorphic operations [13, 28, 29, 61, 62]. In [14], a large-integer multiplier and a Barrett modular reduction are proposed that can accelerate HE operations by 11×.

**GPU-based Acceleration.** GPU is an alternative computing platform to accelerate evaluation functions [6, 21, 24, 42, 48, 59]. Wang et al. [59] have proposed the first GPU acceleration of FHE that targets Gentry-Halevi [34] scheme. Subsequent improvements are reported in [60]. In [58], a GPU-based implementation of BGV scheme [11] is introduced. In [6], a comprehensive study is reported for multi-threaded CPU execution as well as GPU for the BFV scheme. To the best of our knowledge, there is no GPU-accelerated implementation of the CKKS scheme. GPUs normally offer less performance per watt of power than FPGAs by design. Therefore, FPGAs are more suitable candidates for high-performance and low-power secure computation.

**Acceleration of YASHE and LTV Schemes.** Several works [19, 20, 23, 27, 49, 50] focus on improving the performance of YASHE [10] and LTV [45] schemes or their variants. These constructions – based on an overstretched NTRU assumption – are subject to a subfield lattice attack [3] and are no longer secure. In [52], an architecture for YASHE scheme is proposed that provides 25× performance improvement over CPU. However, authors assume unlimited memory bandwidth which renders off-chip memory accesses free of cost and is not a realistic assumption. Pöppelmann et al. [51] have also proposed an architecture for YASHE scheme. Since ciphertexts are prohibitively large to be stored on on-chip memory, authors propose to leverage the idea of Cached-NTT [4, 5] to reduce off-chip memory accesses. In contrast, HEAX relies on the ring isomorphism property and perform independent computation on RNS components. This, in turn, allows us to avoid off-chip memory accesses for small HE parameters and minimize such accesses for large parameters.

## 9 CONCLUSION

In this paper, we introduced a novel set of architectures for Fully Homomorphic Encryption (FHE). To the best of our knowledge, HEAX is the *first* architecture and fully-implemented hardware acceleration for the CKKS FHE scheme. CKKS is the prime candidate for machine learning on encrypted data due to floating-point support of this scheme. The components designed in HEAX can also be used for other lattice-based cryptosystems and other FHE/HE schemes. The proposed architecture provides a unique degree of flexibility that can be readily adjusted for various FPGA chips. As a proof-of-concept, we have implemented HEAX on two different FPGAs with contrasting hardware resources. Moreover, unlike prior FPGA-based acceleration for BFV scheme, our design is not tied to a specific FHE parameter set. We evaluate HEAX on a wide range of FHE parameters demonstrating more than *two orders of magnitude* performance improvements. We hope that HEAX paves the way for large-scale deployment of privacy-preserving computation in clouds.

### Acknowledgments

We would like to thank our shepherd Dr. Timothy Sherwood and our reviewers for their valuable suggestions.